%% file: SnowmassBook-IF02.tex
\documentclass{tcibook}
\usepackage{fancyhea}
\usepackage{work}
\usepackage{bm}       %    enables bold math symbols  e.g.  \bm{\gamma}
\usepackage{graphicx}
\usepackage{multirow}
\usepackage{lineno}
\usepackage{threeparttable}
\usepackage[normalem]{ulem}
\usepackage{color}
\usepackage[colorlinks = true,
            linkcolor = blue,
            urlcolor  = blue,
            citecolor = blue,
            anchorcolor = blue]{hyperref}

\input workshopsymbols.tex     
\def\authorlist#1#2{
    \vskip 0.4in
\begin{center}\begin{large} {\bf  #1 } \end{large}
    \vskip 0.2in
              #2
     \vskip 0.2in
   \end{center}
}

\begin{document}

%%  uncomment this line to use line numbers in drafts:
%\linenumbers

\pagenumbering{roman}

\parindent=0pt
\parskip=8pt
\setlength{\evensidemargin}{0pt}
\setlength{\oddsidemargin}{0pt}
\setlength{\marginparsep}{0.0in}
\setlength{\marginparwidth}{0.0in}
\marginparpush=0pt

% The content begins here

\pagenumbering{arabic}

\renewcommand{\chapname}{chap:intro_}
\renewcommand{\chapterdir}{.}
\renewcommand{\arraystretch}{1.25}
\addtolength{\arraycolsep}{-3pt}

% \thispagestyle{empty}
% \begin{centering}
% \mbox{\null}
% \rightline{\begin{tabular}{l}
% FERMILAB-CONF-xx\\
% SLAC-PUB-xx\\
%  \end{tabular}}
% \vfill

% {\Huge\bf The Future of US Particle Physics}

% \vskip 0.6in

% {\LARGE \bf Report of the 2021  US  Community Study  \\
%      on the Future of Particle Physics

%                   \smallskip

%        organized by the  APS Division of Particles and Fields}

% \vfill

% \input Frontmatter/mainauthorlist.tex

% \vfill

% \end{centering}

% \newpage
% \thispagestyle{empty}

% \mbox{\null}

% \newpage

%\pagenumbering{roman}
% \input Frontmatter/Foreword.tex 

% \newpage
% \thispagestyle{empty}

% \mbox{\null}

% \newpage

% \input Frontmatter/ExecutiveSummary.tex

% \newpage
% \thispagestyle{empty}

% \tableofcontents

% \newpage
%  \pagenumbering{arabic}

% \input Frontmatter/Summary.tex

% \newpage
% \thispagestyle{empty}

% \mbox{\null}

%\input Energy/Energy.tex

% \newpage
% \thispagestyle{empty}
% \mbox{\null}

% \input Instrumentation/Instrumentation.tex

% \newpage
% \thispagestyle{empty}

% \mbox{\null}

% \input Frontmatter/Glossary.tex 

% % \end{document}

%  \newpage
% \thispagestyle{empty}
%  \mbox{\null}

% \input Instrumentation/Header.tex
% \input Instrumentation/IF01/Quantum.tex
 \input Instrumentation/IF02/Photons.tex
% \input Instrumentation/IF03/Tracking.tex
% \input Instrumentation/IF04/TDAQ.tex
% \input Instrumentation/IF05/Gas.tex
% \input Instrumentation/IF06/Calorimetry.tex
% \input Instrumentation/IF07/Electronics.tex
% \input Instrumentation/IF08/Noble.tex
% \input Instrumentation/IF09/Integration.tex
% \input Instrumentation/IF10/Radio.tex

%%%%%%%%%%%%%%%%%%%%%%%%%%%%%%%%%%%%%%%%%%%%%%%%%%

\end{document}

%% file: workshopsymbols.tex
\def\eg{{\it e.g.}}

\def\beq{\begin{equation}}
\def\eeq#1{\label{#1}\end{equation}}
\def\eeqn{\end{equation}}

\newenvironment{Eqnarray}%
   {\arraycolsep 0.14em\begin{eqnarray}}{\end{eqnarray}}
\def\beqa{\begin{Eqnarray}}
\def\eeqa#1{\label{#1}\end{Eqnarray}}
\def\eeqan{\end{Eqnarray}}

\let\bar=\overbar

\def\lsim{\mathrel{\raise.3ex\hbox{$<$\kern-.75em\lower1ex\hbox{$\sim$}}}}
\def\gsim{\mathrel{\raise.3ex\hbox{$>$\kern-.75em\lower1ex\hbox{$\sim$}}}}

\def\del{\partial}
\def\Dslash{\not{\hbox{\kern-4pt $D$}}}
\def\dslash{\not{\hbox{\kern-2pt $\del$}}}
\def\pslash{\not{\hbox{\kern-2pt $p$}}}
\def\ETmiss{\not{\hbox{\kern-4pt $E$}}_T}

\def\Dlr{\mathrel{\raise1.5ex\hbox{$\leftrightarrow$\kern-1em\lower1.5ex\hbox{$D$}}}}

\def\MSB{{\bar{M \kern -2pt S}}}
\def\msb{{\bar{\scriptsize M \kern -1pt S}}}

\def\drb{{\bar{\scriptsize D \kern -1pt R}}}

%% file: Instrumentation/IF02/Photons.tex
\setcounter{chapter}{1} 

%% IMPORTANT:   from this file, refer to the bibliography as   Instrumentation/IF02/bibliography.tex   
%%    refer to a figure   A.pdf  as    Instrumentation/IF02/figures/A.pdf  .

\chapter{Photon Detectors}

\authorlist{C. O. Escobar, J. Estrada, C. Rogan}
   {(contributors from the community)}

\section*{Executive Summary}

The Photon Detectors Topical Group has identified two areas where focused R\&D over the next decade could have a large impact in High Energy Physics Experiments. These areas described here are characterized by the convergence of a compelling scientific need and recent technological advances.

The development of detectors with the capability of counting single photons from IR to UV has been a very active area in the last decade. Demonstration for sensors based on novel semiconductor (CMOS, skipper-CCD, SiPM) and superconducting technologies (MKID, SNSPD and TES) have been performed. These sensors open a new window for HEP experiments in the low photon number regime. Several ongoing and future projects in HEP benefit from these developments (Cosmology, Dark Matter, Neutrinos) which will also have a large impact outside HEP (BES, QIS, Astronomy). The combined scientific needs and technological opportunities make photon counting an ideal area for focused R\&D investment in the coming decade.  R\&D is needed for enabling large arrays of the new sensors, improving their energy resolution, timing, dark counts rates and extending their wavelength coverage. Such investment will secure leadership in photon counting technologies in the US, producing a large impact in HEP, with applications outside HEP. This opportunity is summarized in Ref \cite{PhotonCountingWP}.

A technological solution for the photon detection system in the first two modules of the
DUNE detector exist, based on the by now well-known Arapuca light traps, with wavelength shifters and SiPMs as the photon detector \cite{Arapuca}, \cite{Vertical_drift}. However, new photon detector developments are being considered
for modules 3 and 4. Some of the proposed ideas consist on novel light collectors, the so-called dichroicons which are Winston-style light concentrators made from dichroic mirrors,
allowing photons to be sorted by wavelength, directing the long-wavelength end of broad-band Cherenkov
light to photon sensors that have good sensitivity to those wavelengths, while directing narrow-band shortwavelength
scintillation light to other sensors \cite{Klein}. This technology could be used in water-based liquid scintillators thus realizing a hybrid Cerenkov/scintillator detector. 
Also notable are research and development efforts in new materials that could be directly sensitive to the VUV light, such as amorphous selenium (a-Se) \cite{selenium} and organic semiconductors \cite{organic}. R\&D is needed to move from concept demonstrations to full scale implementations in an HEP experiment. Future investment in the above technologies over the next decade will enhance the science of the DUNE project.

The science for generation, detection and manipulation of light is extremely fast moving and driven mainly from outside our field. The HEP community would benefit from resources dedicated to the implementation of these advanced photonic technologies in particle physics experiments \cite{Young}, \cite{Eftekharian}, \cite{Miyata}.

%The Photon Detectors Topical Group has identified two areas where focused R&D over the next decade could have a large impact in High Energy Physics Experiments. Technological opportunities meet scientific needs in these two areas.

\begin{itemize}
\item[\bf IF02-1]  The development of detectors with the capability of counting single photons from IR to UV has been a very active area in the last decade. We now need to pursue R\&D to 
implement these in HEP experiments by making larger arrays,
improving their energy resolution, timing, dark counts rates and extending their wavelength coverage. 

\item[\bf IF02-2]  New photon detector developments are being considered for planned future neutrino experiments going beyond the current technologies. Concept demonstrations have been done, and we now need to move from a conceptual phase to working detectors.

\end{itemize}

\section{Photon Counting Sensors Enabling HEP}

\begin{figure}
    \centering
    \includegraphics[width=0.5\linewidth]{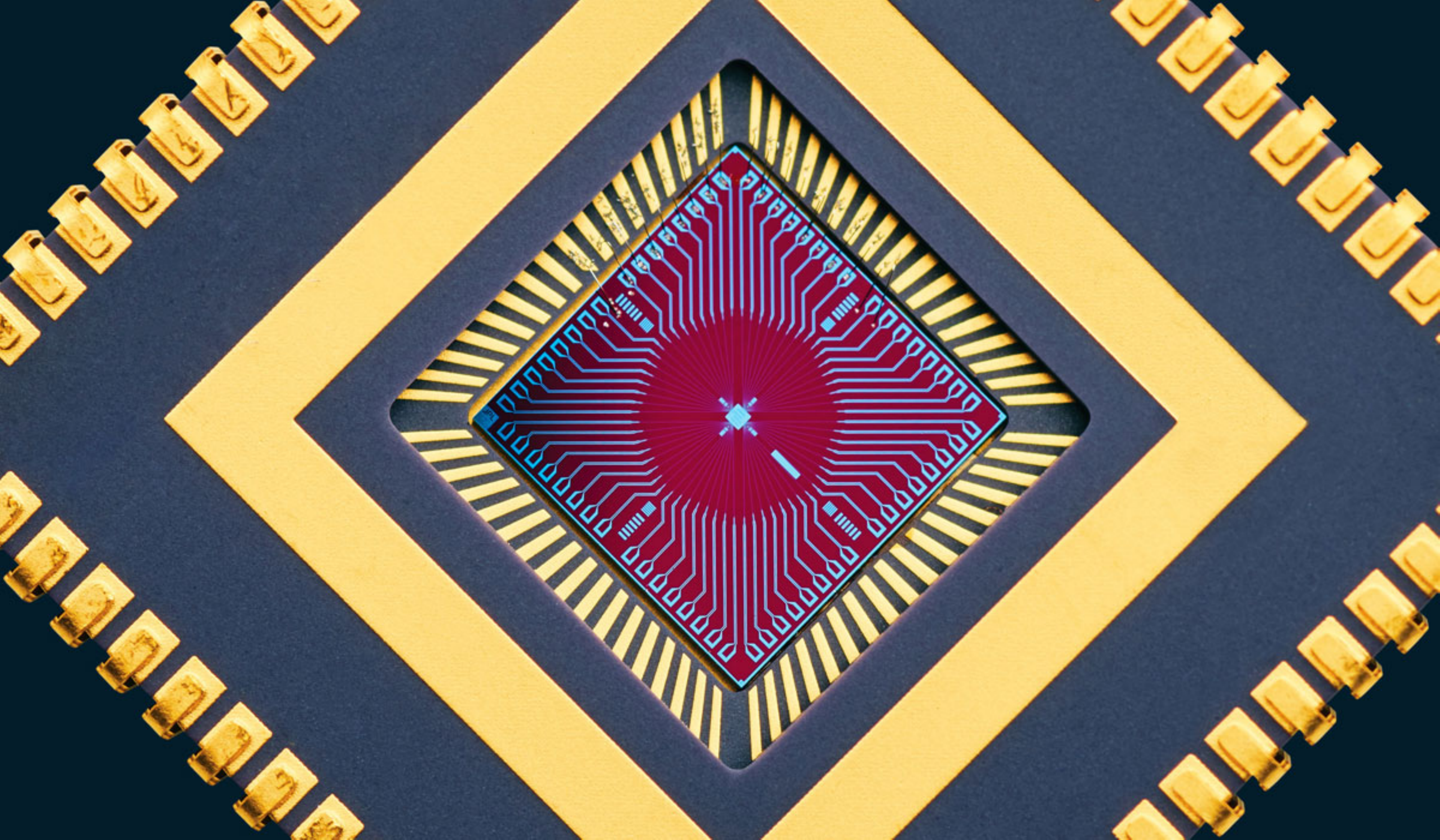}
    \includegraphics[width=0.3\linewidth]{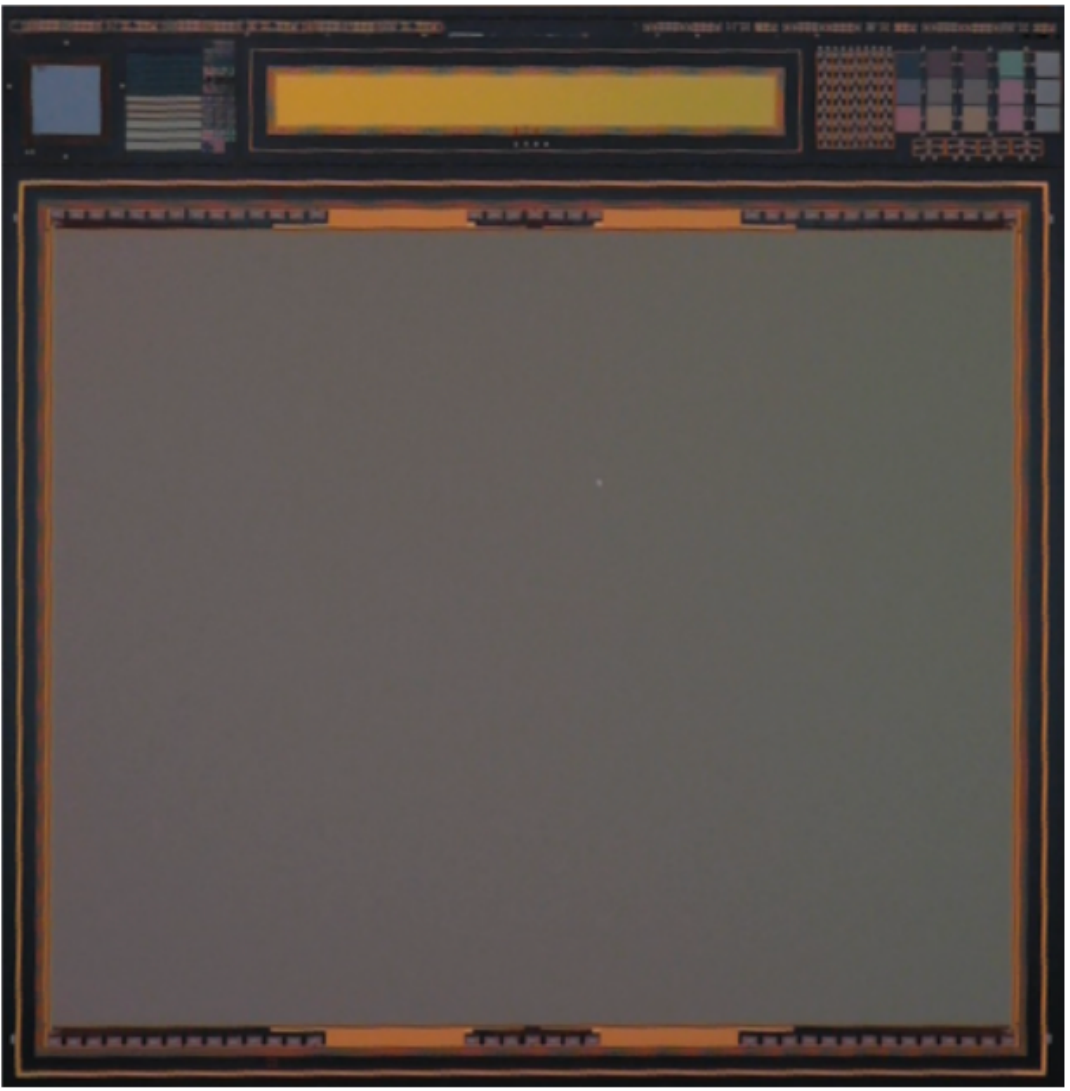}
    \caption{Example recently develop of superconducting and semiconducting photon counting sensors. Left) 64-pixel SNSPD array capable of counting over 1 billion photons per second with time resolution below 100 ps for Astronomy (JPL Microdevices Laboratory). Righ) 1.3 Mpix skipper-CCD capable of deep sub-electron noise (LBNL Micro Systems Laboratory) developed for Astronomy and direct Dark Matter Search. }
    \label{fig:photoncounter}
\end{figure}

\subsection{Science Needs}\label{sec:countingNeeds}

Several novel cosmological facilities for wide-field multi-object spectroscopy were proposed for the Astro2020 decadal review, and are being considered as part of the Snowmass process \cite{MegaMapper}. 
Ground-based spectroscopic observations of faint astronomical sources in the low-signal, low-background regime are currently limited by detector readout noise. Significant gains in survey efficiency can be achieved through reductions by using sensors with readout noise below 1e-. Pushing the current photon counters in the direction of mega pixel arrays with fast frame rate (~10 fps) would make this possible.

Dark matter searches  based on photon counting technologies currently hold the world record sensitivity for low mass electron-recoil dark matter \cite{sensei2019} semiconductors are among the most promising detector technologies for the construction of a large multi-kg experiment for probing electron recoils from sub-GeV DM (skipper-CCDs) \cite{oscura_2020}. Significant R\&D is needed to scale to scale the experiments from the relatively small pathfinders to the multi-kg experiments in the future. 

Single photon counting sensors have also gained importance for their potential as CEvNS detectors. (Coherent Elastic neutrino Nucleus Scattering ) \cite{cevns_2017}. The deposited energy from CEvNS is less than a few keV. Only part of this energy is converted into detectable signal in the sensor (ionization, phonons, etc.) and therefore low threshold technologies are needed. Semiconductor and superconducting technologies with eV and sub-eV energy resolution for photon counting capability in the visible and near-IR are natural candidates to reach the necessary resolution for this application.

Photon counting also enables a wide range of science outside HEP \cite{PhotonCountingWP} including QIS, BES and applications in radiation detection.

\subsection{Technological Opportunities}

Recent advances in photon counting technology addressing the needs in Section \ref{sec:countingNeeds} are discussed here. The advances can be grouped in three areas. 

\subsubsection{Superconducting Sensors}

\paragraph{MKIDs} (microwave kinetic inductance detector) work on the principle that incident photons change the surface impedance of a superconductor through the kinetic inductance effect~\cite{Mattis:1958vo}. The magnitude of the change in surface impedance is proportional to the amount of energy deposited in the superconductor, allowing for single photon spectroscopy on chip. Frequency multiplexed arrays  20,440 pixels with energy resolution R=E/$\Delta$E$\sim$9.5 at 980 nm, and a quantum efficiency of ${\sim}$35\% have been achieved. R\&D focused on larger arrays with higher QE and better energy resolution would address the need discussed in Sec. \ref{sec:countingNeeds}.

\paragraph{SNSPDs} (Superconducting Nanowire Single Photon Detector) is a superconducting film patterned into a wire with nanometer scale dimensions (although recently devices with micrometer-scale widths have been shown to be single-photon sensitive \cite{Korneeva2018}). SNSPDs have been reported with single photon sensitivity for wavelengths out to several microns, timing jitter as low as a few~ps \cite{Korzh:2018}, dark count rates (DCR) down to $6\times10^{-6}$ Hz \cite{Chiles:2021}, and detection efficiency (DE) of 0.98 \cite{Reddy:2020}. They have also been  have been shown to function in magnetic fields of up to 6T \cite{Lawrie:2021}. Current R\&D consist on scaling to large arrays and extending the spectral range for these sensors to address the needs of HEP and Astrophysics.

\paragraph{TES} (Transition-Edge Sensor)  photon detector, which utilizes a patterned superconducting film with a sharp superconducting-to-resistive transition profile as a thermometer, is a thermal detector with a well developed theoretical understanding. 
When a visible or infrared photon is absorbed by a TES, the tiny electromagnetic energy of the photon increases the temperature of the TES and therefore changes its resistance. TES have been developed to measure single photons in quantum communication~\cite{Rosenberg_07, Litaa_08, Fukuda_09, Fujino_11}, for axion-like particle searches ~\cite{Bastidon_16, Eschweiler_15}, direct detection of dark matter particles and astrophysical observations in the wavelengths between ultraviolet and infrared~\cite{Burney_06}.  TES detectors can be multiplexed enabling arrays of large channel counts~\cite{Bender_20, Doriesel_16, Dober_21}. 
Multiplexers for detector arrays using 16,000 TESs have already been successfully implemented~\cite{Bender_20}.  R\&D  exploiting microwave resonance techniques~\cite{Dober_21} have the potential to increase the multiplexing capacity by another factor of 10. 

\subsubsection{Semiconducting Sensors}

\paragraph{skipper-CCDs} Skipper-CCDs have an output readout stage that allows multiple non-destructive sampling of the charge packet in each pixel of the array thanks to its floating gate output sense node. This non-destructive readout has been used to achieve deep sub-electron noise in mega-pixel arrays Skipper CCDs fabricated on high resistivity silicon \cite{Holland_2003} has also demonstrated an extremely low production of dark counts. This technology as motivated to build a new generation of Dark Matter \cite{oscura_2020, PhysRevLett.122.161801} and neutrino experiments \cite{Nasteva_2021}). The R\&D effort here is currently is currently focused on faster readout (10 fps) and large gigapixel arrays.

\paragraph{CMOS} The down scaling of CMOS technology has allowed the implementation of pixels with a very low capacity, and therefore, high sensitivity and low noise (1-2\,$\rm e^-$) at room temperature and high frame rates (50-100\,fps) \cite{ma20154mp}\cite{fowler20105}. Commercial cameras  with  sub-electron noise at 5\,fps are now available. These sensors have not yet played a big role in HEP mainly because of the small pixel size. Active R\&D taking advantage of CMOS fabrication process to address the needs of HEP is ongoing to produce, including the development of new CMOS sensors with non-destructive readout (skipper-CMOS). These sensors could address the readout speed limitations of other semiconductor photon counters. The single photon avalanche diode (SPADs) have also been implemented in standard CMOS technology and integrated with on-chip quenching and recharge circuitry addressing fast timing and radiation tolerance requirements from HEP \cite{APDHep}.

\paragraph{Photon-to-Digital converter} In a PDC, each SPAD is coupled to its own electronic quenching circuit.
This one-to-one coupling provides control on individual SPADs and signals each detected avalanche as a digital signal to a signal processing unit within the PDC.
Hence, PDCs provide a direct photon to digital conversion considering that intrinsically a SPAD is a Boolean detector by design. Digital SiPMs were first reported in 1998 \cite{1998-Aull_SPAD-3D} by the MIT Lincoln Lab and many contributions followed \cite{1998-Aull_SPAD-3D, Aull2016}.
A major step came with microelectronics integration to fabricate both the SPAD and readout quenching circuit in a single commercial process \cite{1997-Zappa_SPAD, 1997-Kindt_SPAD-array, 1998-Kindt_SPAD-CMOS, 1999-Kindt_thesis,  2003-Rochas_thesis, 2003-Rochas_SPAD-CMOS}. 
These innovations led to the first multi-pixel digitally read SPAD arrays \cite{Rochas2DSiPM, 2005-Niclass-Rochas-Charbon_CMOS-SPAD-camera-32x32-800nm}. A recent review can be found in Ref. \cite{Pratte2021_Sensors}.

\subsubsection{Extending wave length coverage}

\paragraph{Ge semiconductor}

Silicon CCDs are commonly utilized for scientific imaging applications in the visible and near infrared. These devices offer numerous advantages described previously, while the skipper CCD~\cite{Tiffenberg2017} adds to these capabilities by enabling multiple samples during readout to reduce read noise to negligible levels~\cite{AguilarPRL2017,sensei2018}. CCDs built on bulk germanium offer all of the advantages of silicon CCDs while covering an even broader spectral range. The R\&D in this area will extend the photon counting capabilities of semiconductor into the IR.

\paragraph{UV}

Jet Propulsion Laboratory showed that CCD sensitivity can be increased closed to the reflection-limited quantum efficiency of silicon down \cite{Hoenk1992_UVCCD_Delta}.
This was done by blocking the surface fields and traps through the epitaxial growth of a strongly doped very thin silicon layer (delta-doping).
Quantum efficiency exceeding 50\% were demonstrated in CCDs down to 125~nm wavelength \cite{nikzad2012delta}.
The method was demonstrated efficient on backside illuminated SPAD based detectors by Schuette in~2011\cite{schuette2011mbe}.
Other methods to address the surface fields and traps issues were also demonstrated \cite{nanver2014robust}.
Work is being done at Caltech (D.~Hitlin) to enhance SiPMs for the detection of the fast scintillation component of BaF$_2$ \cite{hitlin2021progress}.
An extensive study of the delta-doping approach to enhance VUV sensitivity in frontside illuminated SPAD based detectors was done by Vachon \cite{Vachon20221-MScA}.

\section{Photon Detectors For Neutrino Experiments}

\begin{figure}
    \centering
    \includegraphics[width=1.\linewidth]{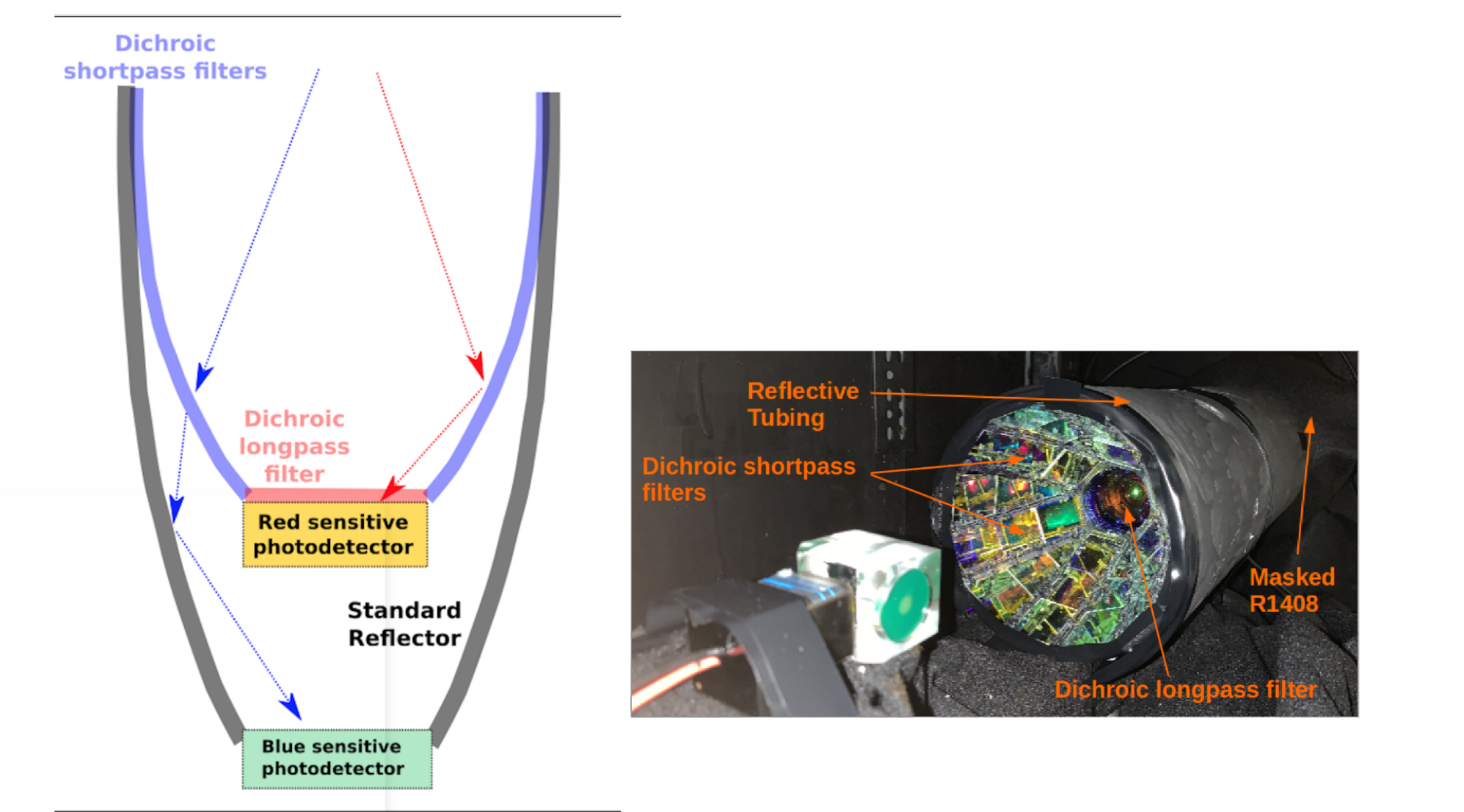}
    \caption{Example of photon detector development for neutrinos:
    the dichroicon, from arXiv:2203.07479}
    \label{fig:pdNeutrinos}
\end{figure}

A large number of outstanding questions remain to the fundamental nature of the neutrino, which can be probed through the use of higher energy ( $\mathcal{O}$(MeV) $< E < \mathcal{O}$(GeV)) neutrino sources (\eg, accelerator and atmospheric neutrinos). The nature of these remaining puzzles break into the distance over which the neutrinos are allowed to propogate before being detected. Thus the future class of experiements are classified as ``short-baseline'' and ``long-baseline'' experiments. 

The next generation long-baseline neutrino experiments aim to answer the questions of the exact ordering of the neutrino mass states, known as the mass hierarchy, as well as the size of the CP-violating phase $\delta$. These, as yet unknown quantities, remain one of the last major pieces of the Standard Model of particle physics and offer the opportunity to answer such fundamental questions as ``what is the origin of the matter/antimatter asymmetry in the universe?'' and ``do we understand the fundamental symmetries of the universe?''. By measuring the asymmetry between appearance of electron neutrinos from a beam of muon neutrinos ($P(\nu_{\mu} \rightarrow \nu_{e}$)) compared to the appearance of electron antineutrinos from a beam of muon antineutrinos and $P(\bar{\nu}_{\mu} \rightarrow \bar{\nu}_{e}$)) as well as the precise measurement of the $\nu_{e}$ energy spectrum measured at the far detector, both the CP violating phase ($\delta_{CP}$) and the mass hierarchy can be measured in the same experiment.

The Short-Baseline Neutrino (SBN) program aims to address the anomalous neutrino results seen by the LSND and MiniBooNE which suggest the possible existence of a eV mass-scale sterile neutrino. However, the experimental landscape is perplexing since a number of other experiments utilizing a range of different neutrino sources which should have been sensitive to such a sterile neutrino have observed only the standard three neutrino oscillations. In this landscape, the conclusive assessment of the experimental hints of sterile neutrinos becomes a very high priority for the field of neutrino physics.

To address both of these areas of neutrino research, large scale noble element time projection chambers (TPC's) \cite{Radeka1974, Rubbia1977} play a central role and offer an opportunity to perform discovery level measurements through the enhancement of their capabilities. In a noble element TPC, particles interact with the medium and deposit their energy into three main channels: heat, ionization, and scintillation light. Depending on the physics of interest, noble element detectors attempt to exploit one or more of these signal components. Liquid Noble TPC's produce ionization electrons and scintillation photons as charged particles traverse the bulk material. An external electric field allows the ionization electrons to drift towards the anode of the detector and be collected on charge sensitive readout or transform energy carried by the charge into a secondary pulse of scintillation light. 

A technological solution for the photon detection system in the first two modules of the
DUNE detector exist, based on the by now well-known Arapuca light traps, with wavelength shifters and SiPMs as the photon detector \cite{Arapuca}, \cite{Vertical_drift}. This approach is also being used in other short baseline neutrino oscillation experiments\cite{SBND}, but not exclusively \cite{icarus}. Beyond this,new photon detector developments are being considered for modules 3 and 4 as well as for other approved or proposed experiments\cite{Klein}. 

\paragraph{Going beyond DUNE's first two modules}

Quoting from the executive summary of the Snowmass IF2 White Paper\emph{Future Advances in Photon-Based Neutrino Detectors}\cite{Klein},
"large-scale, monolithic detectors that use either Cherenkov or scintillation light have played major roles in nearly every discovery of neutrino oscillation phenomena or observation of
astrophysical neutrinos. New detectors at even larger scales are being built right now, including JUNO \cite{JUNO},
Hyper-Kamiokande \cite{HyperK}, and DUNE \cite{Dune}.

These new technologies will lead to neutrino physics and astrophysics programs of
great breadth: from high-precision accelerator neutrino oscillation measurements, to detection of reactor and solar
neutrinos, and even to neutrinoless double beta decay measurements that will probe the normal hierarchy regime.
They will also be valuable for neutrino applications, such as non-proliferation via reactor monitoring".

"Of particular community interest is the development of hybrid Cherenkov/scintillation detectors, which can simultaneously
exploit the advantages of Cherenkov light|reconstruction of direction and related high-energy particle identification (PID)
and the advantages of scintillation light, high light-yield, low-threshold detection with low-energy PID. Hybrid
Cherenkov/scintillation detectors could have an exceptionally broad dynamic range in a single experiment, allowing
them to have both high-energy, accelerator-based sensitivity while also achieving a broad low-energy neutrino
physics and astrophysics program. Recently the Borexino Collaboration \cite{Borexino} has published results showing that even in a detector with standard scintillator and no special photon sensing or collecting, Cherenkov and scintillation light can be discriminated well enough on a statistical basis that a sub-MeV solar neutrino direction peak can be seen. Thus the era of hybrid detectors has begun, and many of the enabling technologies described here will make full
event-by-event direction reconstruction in such detectors possible".

Among the new technologies of relevance for this topical group, it should be mentioned 
\paragraph{New Photon Sensors}: New advances in the science of photomultiplier tubes, including long-wavelength
sensitivity, and significant improvements in timing even with devices as large as 8 inches, make hybrid
Cherenkov/scintillation detectors even better, with high light yields for both Cherenkov and scintillation light
with good separation between the two types of light. Large Area Picosecond Photon
Detectors (LAPPDs) \cite{LAPPD} have pushed photon timing into the picosecond regime, allowing Cherenkov/scintillation
separation to be done even with standard scintillation time proles. The fast timing of LAPPDs also makes
reconstruction of event detailed enough to track particles with the produced photons.

\paragraph{New Photon Collectors}: Dichroicons, which are Winston-style light concentrators made from dichroic mirrors,
allow photons to be sorted by wavelength thus directing the long-wavelength end of broad-band Cherenkov
light to photon sensors that have good sensitivity to those wavelengths, while directing narrow-band shortwavelength
scintillation light to other sensors. Dichroicons are particularly useful in high-coverage hybrid
Cherenkov/scintillation detectors.

%\paragraph{point to some examples with references}

%%%%%%%%%%%%%%%%%%%%%%%%%%%%%%%%%%%%%%%%%%

%  If you would like to use BibTEX for the bibliography, please feel free to do so.  It is not required.

%  To use BibTeX,

%    1.  uncomment the following two lines,
%    2.  comment out everything below from  \begin{thebibliography}{99}   to \end{thebibliography).
%    3.  create the file  myreferences.bib in this directory, and process this file in the usual way

\bibliographystyle{JHEP}
\bibliography{Instrumentation/IF02/IF02} 

%%%%%%%%%%%%%%%%%%%%%%%%%%%%%%%%%%%%%%%%%

%\begin{thebibliography}{99}

%\input Instrumentation/IF02/bibliography.tex

%\end{thebibliography}